\documentclass[aip,rsi,reprint,numerical,floatfix]{revtex4-1}
\usepackage{graphicx}
\usepackage{hyperref}
\usepackage{xcolor}

\usepackage[utf8]{inputenc}
\usepackage[T1]{fontenc}
\usepackage{mathptmx}

\pdfoutput=1

\begin{document}
\pdfsuppresswarningpagegroup=1

\title{Piezoelectric-based uniaxial pressure cell with integrated force and displacement sensors}

\author{Mark E. Barber}
\email{barber@cpfs.mpg.de}
\author{Alexander Steppke}
\affiliation{Max Planck Institute for Chemical Physics of Solids, N\"{o}thnitzer Stra\ss e 40, Dresden 01187, Germany}
\author{Andrew P. Mackenzie}
\affiliation{Max Planck Institute for Chemical Physics of Solids, N\"{o}thnitzer Stra\ss e 40, Dresden 01187, Germany}
\affiliation{Scottish Universities Physics Alliance (SUPA), School of Physics and Astronomy, University of St.
Andrews, North Haugh, St. Andrews KY16 9SS, United Kingdom}
\author{Clifford W. Hicks}
\email{hicks@cpfs.mpg.de}
\affiliation{Max Planck Institute for Chemical Physics of Solids, N\"{o}thnitzer Stra\ss e 40, Dresden 01187, Germany}

\date{\today}

\begin{abstract}

We present a design for a piezoelectric-driven uniaxial stress cell suitable for use at ambient and cryogenic temperatures, and that incorporates both a displacement and a force sensor.
The cell has a diameter of 46~mm and a height of 13~mm.
It can apply a zero-load displacement of up to $\sim$45~$\mu$m, and a zero-displacement force of up to $\sim$245~N.
With combined knowledge of the displacement and force applied to the sample, it can quickly be determined whether the sample and its mounts remain within their elastic limits.
In tests on the oxide metal Sr$_2$RuO$_4$, we found that at room temperature serious plastic deformation of the sample onset at a uniaxial stress of $\sim$0.2~GPa, while at 5~K the sample deformation remained elastic up to almost 2~GPa.
This result highlights the usefulness of in situ tuning, in which the force can be applied after cooling samples to cryogenic temperatures.
\end{abstract}

\maketitle

\section{Introduction}

Uniaxial pressure can be a powerful probe and tuning parameter for the electronic properties of materials.
It is a fundamentally different probe from hydrostatic pressure, because it can directly lift the rotational symmetry of a crystal lattice.
For example, while hydrostatic pressure causes a slow decrease in the critical temperature of the superconductor Sr$_2$RuO$_4$~\cite{Shirakawa1997}, uniaxial pressure drives a rapid increase~\cite{Hicks2014,Taniguchi2015}.

One way to apply uniaxial stress is with mechanical springs, or, for in situ tunability, gas-filled bellows.
These mechanisms can provide large forces at a low spring constant.
This makes the force on the sample a well-controlled variable, because it is essentially independent of small displacements, due, for example, to differential thermal contraction, or to deformation of the sample or its mounts under the applied load.
It is appealing however to use piezoelectric actuators for in situ tunability, because they are mechanically simple, and can generate large forces in a small volume.
Piezoelectric actuators have high spring constants.
Although they can apply sufficient force to compress solid samples by $\sim$1\% or more, this force is not automatically independent of micron-scale deformations that the sample or apparatus might undergo in response.

A uniaxial stress cell based on piezoelectric actuators was presented by some of the present authors in 2014~\cite{Hicks2014RSI}.
The actuators are extension actuators, the most common type of actuator.
To cancel their own thermal contraction and allow operation over a wide temperature range, including cryogenic temperatures, the actuators are arranged into two identical sets, placed so that their action on the sample is of opposite sign: extension of the ``compression'' actuators compresses the sample, and of the ``tension'' actuators tensions the sample.
The basic concept of these cells is to take advantage of the high spring constant of the actuators to build cells whose spring constants greatly exceed those of typical samples.
In this way, the displacement applied to the sample, rather than the force, becomes the well-controlled variable, because it is unaffected by the force generated by the sample in response.
The versatility of this configuration has been demonstrated in a number of different measurements.
The original cell and updated versions have been used for measurements including magnetic susceptibility~\cite{Steppke2017}, resistivity~\cite{Barber2018,Stern2017,Brodsky2017}, X-ray scattering~\cite{Zheng2018,Kim2018}, NMR~\cite{Kissikov2017,Kissikov2018}, and scanned probe microscopy~\cite{Watson2018}.  

Piezoelectric actuators are hysterestic, especially over wide temperature and voltage ranges, so a separate sensor of the applied displacement is needed. 
In most of the cells used in the above-referenced experiments this has been a capacitive sensor placed in parallel with the sample.
The sample strain is determined, essentially, as the applied displacement divided by the strained length of the sample.

In practice, this approach has limitations.
One is that in practice the cell's spring constant is not always much larger than that of the sample, such that deformation of the cell in response to the load from the sample may introduce systematic error into the displacement sensor reading.
Cells of diameter 25--50~mm and height $\sim$15~mm, designed to fit into typical cryostats, typically have spring constants of $\sim$$10^7$~N/m.
If the strained length of the sample is 2~mm and its Young's modulus is 100~GPa (a typical value for a metal) its cross-section is limited to 0.2~mm$^2$ to keep the sample spring constant below the cell spring constant.
Ideally its cross-section should be much less than this. 

In addition, the sample is typically held in place with epoxy, and this epoxy will also deform under the applied load, further eroding the accuracy of the displacement sensor as a sensor of the state of the sample.
The precise mounting achieved varies from sample to sample, making accurate comparison of the strains in different samples very difficult.
Finally, the sample might relieve the applied strain through non-elastic deformation, and with a displacement sensor alone one must often repeat measurements to check if the sample is still in its elastic regime.

Here, we present a piezoelectric-based uniaxial stress cell with an integrated sensor that measures the force on the sample, eliminating the need to try to infer the applied force from the applied displacement.
The cell also still has a displacement sensor, and combined knowledge of force and displacement allows rapid detection of non-elastic deformation in either the sample or the sample mounts.
The main challenge in implementing the force sensor is geometric: it must be placed in series with the sample, which in a compact space is a more difficult arrangement than a displacement sensor in parallel.
The second challenge is that the spring constant of the force sensor must be large, so that it does not absorb too much of the limited displacement generated by the actuators.

\section{Design of the new pressure cell}

Various other configurations have been published in which piezoelectric actuators are used to apply uniaxial stress.
Bending actuators~\cite{Stillwell1968, Overcash1981} and shear actuators~\cite{Gannon2015} have been applied.
However both actuator types have an unfortunate trade-off: if they are made longer, to obtain larger displacements, then their spring constant against bending and correspondingly the maximum force they can apply falls rapidly.
Thin samples can be strained by direct attachment to piezoelectric actuators~\cite{Shayegan2003,Chu2012}, but the achievable strain range is limited to that of the actuator: $\sim$$10^{-3}$.

\begin{figure}[ptb]
\includegraphics{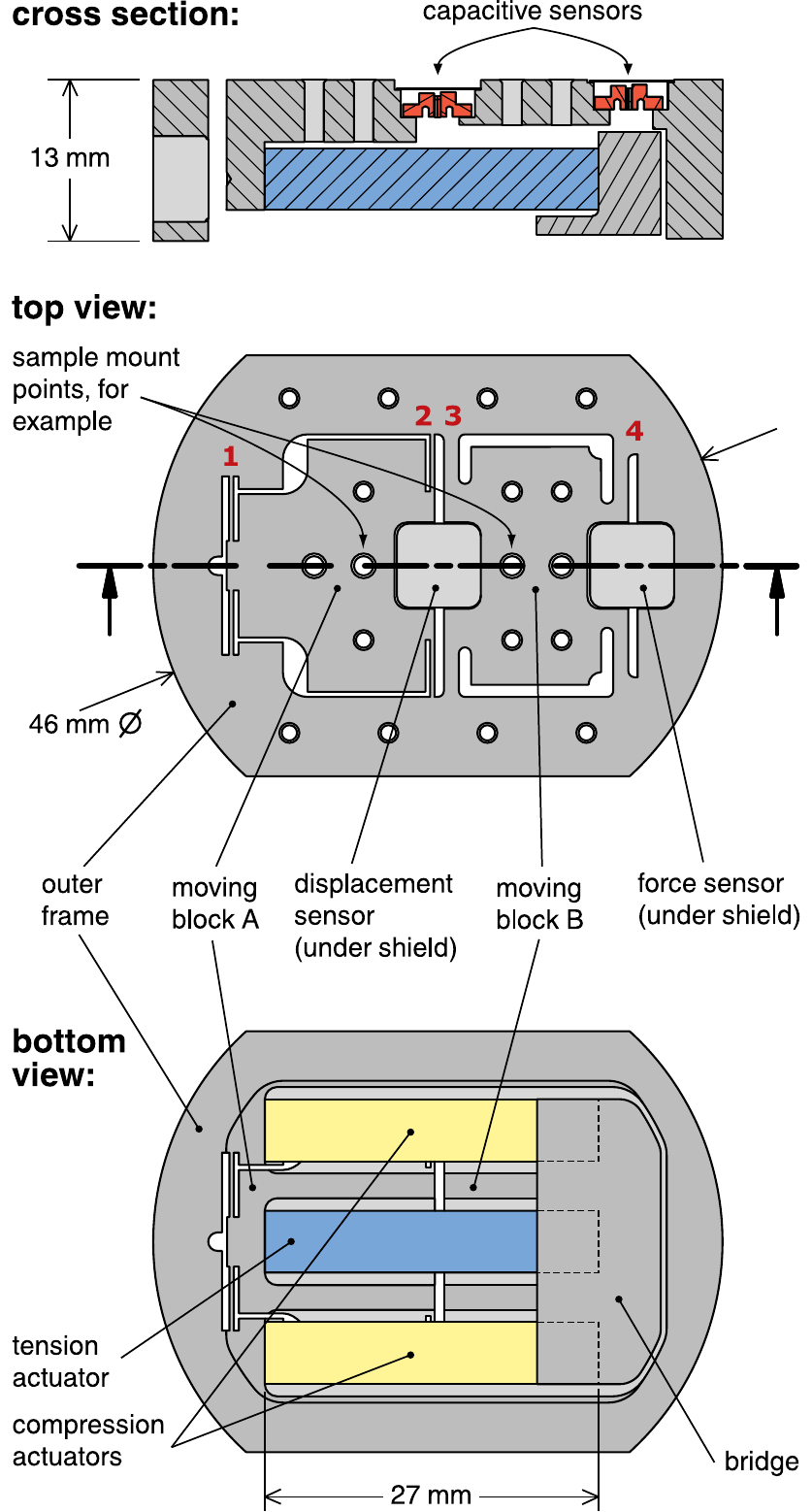}
\caption{\label{1605A}A uniaxial pressure cell with both displacement and force sensors.
Piezoelectric actuators drive motion of moving block A.
Force applied to the sample by this motion is transferred to block B, which moves slightly in response.
The flexures guiding the moving blocks are labelled 1--4; their dimensions are: 1: 0.35 $\times$ 5.2 $\times$ 10.5~mm; 2: 0.35 $\times$ 4.6 $\times$ 5.0~mm; 3: 1.2 $\times$ 4.0 $\times$ 5.0~mm; and 4: 1.2 $\times$ 4.0 $\times$ 4.0~mm.}
\end{figure}

Our new design is presented in Fig.~\ref{1605A}.
It is composed of an outer frame and two moving blocks, labelled A and B.
The moving blocks are joined to the frame through flexures that guide their motion to be in a line along the axis of the cell.
The actuators drive motion of block A, and for compactness they are placed underneath it.
The flexures guiding block A are thin, so that block A has a low resistance to motion.
Placing the actuators underneath introduces large torques: the actuators apply force to the bottom portion of block A, while the counterforce from the sample is applied at the top of block A.
The flexures guiding block A resist this torque, and in the process provide mechanical coupling between the actuators and sample.

The arrangement of actuators to cancel their thermal contraction can be seen in the bottom view.
The bridge joins the tension and compression actuators, but is not joined to the outer frame.
The actuators \emph{lengthen} along their poling direction as they are cooled \cite{PICatalog}.
This expansion pushes the bridge rightward in the figure, but because the expansion of the compression and tension actuators is ideally the same no net displacement is applied to block A.

The sample is mounted between blocks A and B, using whichever combination of attachment points is most convenient.
The displacement and force sensors are both capacitive sensors: by measuring $C$ in $C = \epsilon_0 A/d$, $d$ is determined.
Their functional difference is due to their placement.
The displacement sensor measures the displacement applied between blocks A and B.
The force sensor measures the displacement between block B and the frame: force applied to the sample is transferred to block B, which moves slightly in response.
With knowledge of the spring constant of the flexures that guide block B, the force on the sample can be determined.
The force sensor is intended to be the primary sensor, however, as noted above, the displacement sensor remains useful for diagnostic purposes.

We now estimate the performance limits of this pressure cell.
The application of this cell is scientific measurement, for which maximum performance will generally be of greater interest than long fatigue life.
We therefore take limits at the upper end of the performance of the actuators and material of the cell body.
In practice the limits one is prepared to explore will depend on the importance of the measurement underway.
The discussion here is specific to this particular design, but illustrates the process and shows approximately what is achievable for a cell of this size.
We consider the limits of the cell in three stages.
(1) Limits set by the performance of the piezoelectric actuators and the stiffness of the cell.
(2) Limits imposed by the elastic limit of the material of the body of the cell.
(3) Limits imposed by the tensile strength of the actuators.

\textit{1. Performance of the actuators, and stiffness of the cell.}
The performance of the cell may be specified with two numbers: the maximum displacement that can be applied to a sample of zero spring constant, $d_{\mathrm{max},0}$, and the maximum force that can be applied to a sample of infinite spring constant, $F_{\mathrm{max},0}$.
$d_{\mathrm{max},0}$ is the displacement generated by the actuators within the voltage limits one is prepared to explore.
$F_{\mathrm{max},0} = d_{\mathrm{max},0} \times k_{\mathrm{cell}}$, where $k_{\mathrm{cell}}$ is the spring constant of the cell: if the sample is infinitely stiff, the displacement generated by the actuators deforms the cell rather than the sample.

\begin{figure}[ptb]
\includegraphics{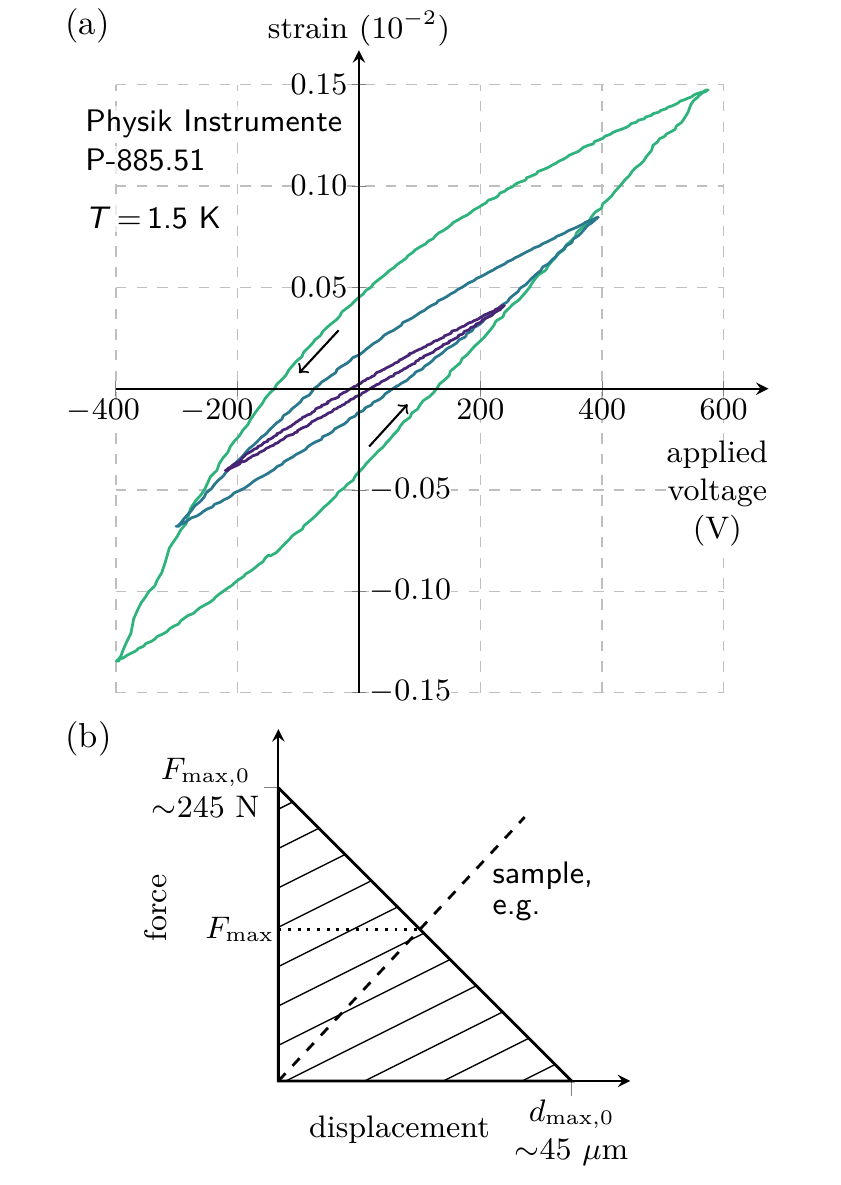}
\caption{\label{piezo}(a) Strain generated in an 18 mm-long piezoelectric actuator~\cite{PICMA} at 1.5 K over voltage sweeps with three different ranges.
The actuator was held under $\sim$1~N of compression.
The displacement generated by the actuator was measured with an optical interferometer.
The curves are each placed along the $y$-axis to be centred on zero strain at $V = 0$.
(b) The approximate limits on displacement and force this cell can apply are indicated by the hatched region.
These limits are set by the limits of the actuators and the elastic limit of the material of the cell body.
An example of a sample with a linear force-displacement relationship is also shown, along with the maximum force that could be applied to this sample, $F_\mathrm{max}$.
}
\end{figure}

The actuators are Physik Instrumente PICMA\textsuperscript{\textregistered} piezoelectric actuators.
In Fig.~\ref{piezo}(a) we show the strain generated in a nearly-free piezoelectric actuator~\cite{PICMA} at 1.5~K as a function of applied voltage, for three voltage ranges.
At cryogenic temperatures we have found $-$300 to $+$400~V to be a safe voltage range that does not lead to obvious degradation of the actuators.
At $-$300~V and $+$400~V the actuator strain is $-7 \cdot 10^{-4}$ and $8 \cdot 10^{-4}$, respectively.
The actuators in this cell are 27~mm in length, so with strains of $-7 \cdot 10^{-4}$ in the tension and $8 \cdot 10^{-4}$ in the compression actuators the generated displacement is 40~$\mu$m.

We calculate approximately the spring constant of the cell, and then check with measurement.
Where finite element analysis (FEA) is required we take the Young's modulus of the titanium to be 103~GPa.
In the FEA it is also necessary to specify the location of the sample, as this determines the torques that blocks A and B experience.
We take the sample to be 0.5~mm above the upper surface of the cell.
(An illustration of a mounted sample is shown in Fig.~\ref{MountedSample}(a).)
Our calculation considers three contributions to the deformation of the cell.
\begin{enumerate}
\item The Young's modulus of the actuators is approximately 40~GPa\cite{PIUserManual}, so the combined spring constant of the set of three actuators is $\approx$25~N/$\mu$m.  
\item As noted above, block A experiences a large torque, which is resisted by the flexures but not with infinite stiffness.
Based on FEA of the deformation of the flexures (and neglecting any contribution to rotational stiffness from the actuators), the spring constant for rotation of block A, as seen at the position of the sample, is 27 N/$\mu$m.
\item Block B is intended to move under the applied load from the sample, and its spring constant is a design decision: a lower spring constant improves the sensitivity of the force sensor and reduces the relative effects of thermal drift within the sensor, but also reduces the total spring constant of the apparatus.
We selected a spring constant for block B of 20~N/$\mu$m.
\end{enumerate}

These spring constants are combined in series, yielding $k_\mathrm{cell} = 7.9$~N/$\mu$m, and $F_\mathrm{max,0} = 316$~N.
The spring constant of the cell was measured at room temperature using a laser interferometer; see the appendix for a photograph of the setup.
The measured spring constant was 7.8~N/$\mu$m.
We note that at higher temperatures the actuators respond more strongly to applied voltage and can be driven further.
However, as we now describe, the limits set by the material of the pressure cell are similar -- $d_\mathrm{max,0} \sim 45$~$\mu$m and $F_\mathrm{max,0} \sim 245$~N -- and these will not increase if $T$ is raised.

\textit{2. Elastic limit of the material of the pressure cell.}
The cell is made from titanium, and the yield stress of titanium varies depending on grade~\cite{Leyens2003}.
We used grade 2 titanium, and we choose a maximum stress within the material of the cell of 300~MPa.
At room temperature, this is relatively high and not suitable for long fatigue life, however the elastic limit will increase as $T$ is reduced. 

FEA of displacement of block A with no load from the sample indicates that the maximum stress reaches 300~MPa when it is displaced by $\sim$45~$\mu$m; details of the FEA are given in the appendix.
This maximum stress occurs within the roots of the flexures guiding block A, and so it is approximate: the FEA results depend strongly on the exact fillet radius at the root of the flexures.
Nevertheless, this simulation indicates that the elastic limits of the flexures on block A approximately match the performance limit of the actuators.

To simulate an infinite-spring-constant sample, we lock the displacement of block A to that of block B at the position of the sample.
Block A will still rotate under the torque it experiences, and we again suppose that the only resistance to this torque are the flexures that guide block A.
The 300 MPa limit is reached, again within the roots of the flexures, at a sample load of $\sim$245~N.
Therefore, we take $F_{\mathrm{max},0} \sim 245$~N.

\textit{3. Force limit on the actuators.}
The actuators are designed to handle compression but not tension; the manufacturer specifies a maximum compressive load on the actuators of 30~MPa (corresponding to 750~N for the $5 \times 5$ mm cross-section actuators used in this cell) and a maximum tensile load of $\sim$10\% of this~\cite{PIUserManual}.
However in this cell design, when the sample is compressed the tension actuator comes under a tensile load at least as large as the applied force.
(Conversely, when the sample is tensioned, the compression actuators come under tension.
However because in this case the tensile load is split between two actuators, the problem is less severe.) 

Here, we demonstrate operation of the cell up to a maximum compressive force of 75~N.
We have separately tested the mechanical strength of one actuator by suspending weight from it at room temperature; it broke apart at a 285~N tensile load.
The general solution when piezoelectric actuators must carry tensile loads is to pair the actuators with pre-load springs, such that the actuator itself always remains under a compressive load.
For compactness, the present design does not incorporate pre-load springs; we take the chance that the actuators can occasionally be used outside their specified tensile load limit, at the cost of potentially reduced lifetime.

The performance limits of the device are summarised in Fig.~\ref{piezo}(b).
The two parameters $d_{\mathrm{max},0}$ and $F_{\mathrm{max},0}$ delineate an approximate safe region of displacement and force over which the cell can be operated, both in compression and in tension.
For a sample of constant spring constant $k$ (that is, whose force-displacement curve is linear), the maximum force that can be applied is approximately
\[
F_\mathrm{max} = F_{\mathrm{max},0} \left(1 + \frac{F_{\mathrm{max},0}}{d_{\mathrm{max},0} k} \right)^{-1}.
\]
It should be stressed that $k$ is the spring constant of everything between the attachment points on blocks A and B: not only the sample itself, but also any plates and epoxy used to hold the sample.

We finish this section with some details on the capacitive sensors.
For a maximum displacement of 45~$\mu$m a sensible initial plate spacing for the displacement sensor is $\sim$50~$\mu$m.
For the force sensor, the maximum force of 245~N on the 20~N/$\mu$m flexures gives a displacement of 12.3~$\mu$m, so a sensible initial plate spacing is $\sim$20~$\mu$m.
In this design the area of the plates in both sensors is $\sim$3.9~mm$^2$, giving an initial capacitance $C = \epsilon_0 A/d$ of $\sim$0.7~pF and $\sim$1.7~pF for the displacement and force sensors respectively.
A precision capacitance bridge can detect changes in capacitance on the order of 10$^{-5}$~pF, yielding noise-limited sensitivities on the displacement and force capacitances of $\sim$0.5~nm and 5~mN, respectively.

\section{Testing and Results}

\begin{figure}[pt]
\includegraphics{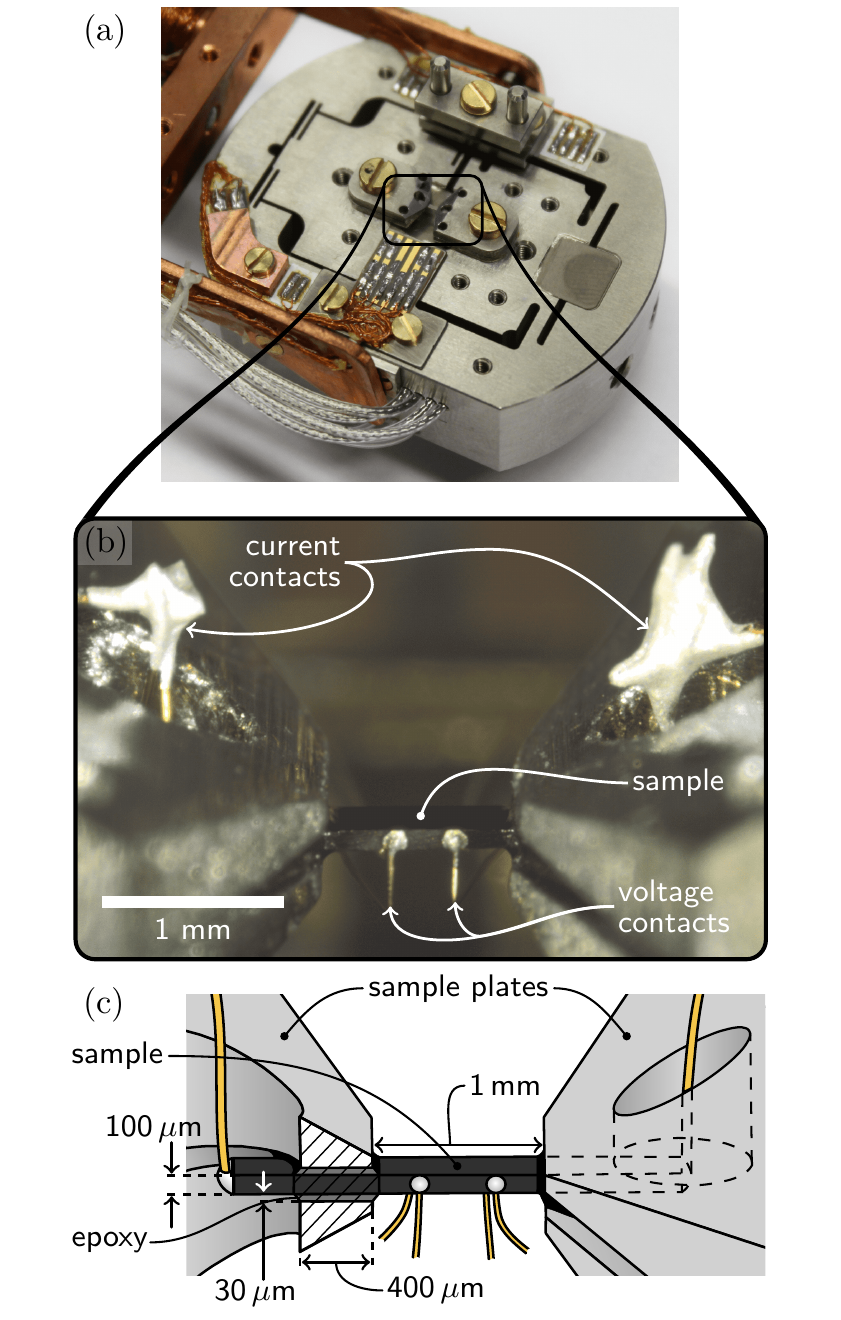}
\caption{\label{MountedSample}(a) Assembled pressure device.
(b) and (c) The mounted sample used for the measurements presented in Fig.~\ref{FigLT}.}
\end{figure}

We now show test results.
All the test samples were mounted in the device using epoxy (Stycast\textsuperscript{\textregistered} 2850FT) sandwiched between two sample plates at each end of the sample, as shown in Fig.~\ref{MountedSample}.
A thorough analysis of this mounting method has been described elsewhere~\cite{Hicks2014RSI,Barber2017}, but the salient features are as follows.
The exposed portion of the sample needs to be long enough to incorporate any voltage contacts, or other measurements apparatus, with some margin to ensure good strain homogeneity within the measured region.
The thickness of the sample is then chosen such that the buckling strain limit of the sample exceeds the desired maximum compressive strain.
The epoxy thickness is a trade-off between two tendencies.
If the epoxy layer is thicker, stress concentration within the ends of the sample and the shear stress within the epoxy is reduced.
However, the load transfer length $\lambda$, the length scale over which force is transferred from the sample plates to the sample, also increases.
$\lambda \sim \sqrt{Etd/2G}$, where $E$ is the sample's Young's modulus, $t$ the sample thickness, $d$ the epoxy thickness and $G$ the epoxy's shear modulus.
Here, as shown in Fig.~\ref{MountedSample}(c), the overlap between the sample and sample plates is 400~$\mu$m, so we target $\lambda \sim$~200~$\mu$m.
Estimating $G$~=~6~GPa\footnote{The elastic properties of Stycast 2850FT with Catalyst 23LV have not been measured at very low temperatures.
By comparing the measured Young's modulus with those from other Catalysts that have been extended to lower temperatures and with the unfilled version of Stycast, 1266, we estimate a Young's modulus of 15~GPa for Stycast 2850FT with Catalyst 23LV at low temperatures.
The shear modulus of an isotropic material is $G = E / 2(1 + \nu)$, where $\nu$ is Poisson's ratio.
We take $\nu \sim 0.3$, yielding $G \sim 6$~GPa.
Further details are given in Ref.~\onlinecite{Hicks2014RSI}} for Stycast 2850FT at cryogenic temperatures, $E$~=~176~GPa for Sr$_2$RuO$_4$\cite{Paglione2002}, and $t$~=~100~$\mu$m yields $d \sim$ 30~$\mu$m.

The displacement-capacitance relations of the sensors were obtained using a laser interferometer, in air, in vacuum at room temperature, and at cryogenic temperatures.
The spring constant of block B was measured by hanging known weights from the block and measuring the deflection.
A value of 19~N/$\mu$m, close to the target value, was obtained.
Further details of these measurements are given in the appendix.

We now report measurements on test samples.
Temperature and humidity variation were found to lead to unwanted drift in the measured capacitances during measurements in air, so all measurements reported here were done under vacuum in a cryostat with active temperature control.
By heating slightly above room temperature we could attain a temperature stability of better than 10~mK at 298~K, and by heating against the cooling power of a 1K pot we could hold the temperature at 5~K with a stability of $\sim$5~mK.

\begin{figure}[pt]
\includegraphics{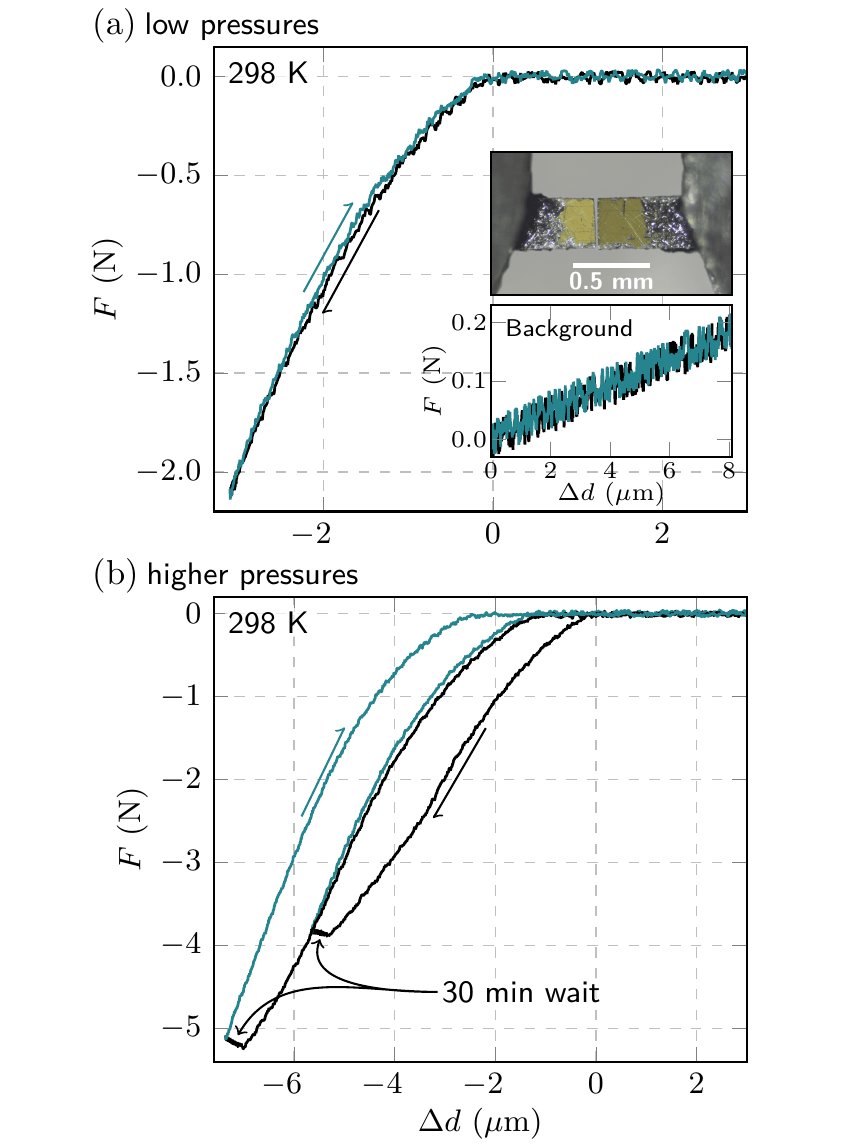}
\caption{\label{FigRT_Ti}Force-displacement curves at room temperature for a split titanium sample at (a) forces up to $\sim$2~N, and (b) higher forces.
A photograph of the sample is shown in an inset of panel (a).
The second inset shows the background coupling between the force sensor and applied displacement, measured with the two sample halves out of contact and so with a true applied force of zero.
This background is subtracted off from the data plotted in the main panels.}
\end{figure}

\begin{figure*}[tp]
\includegraphics{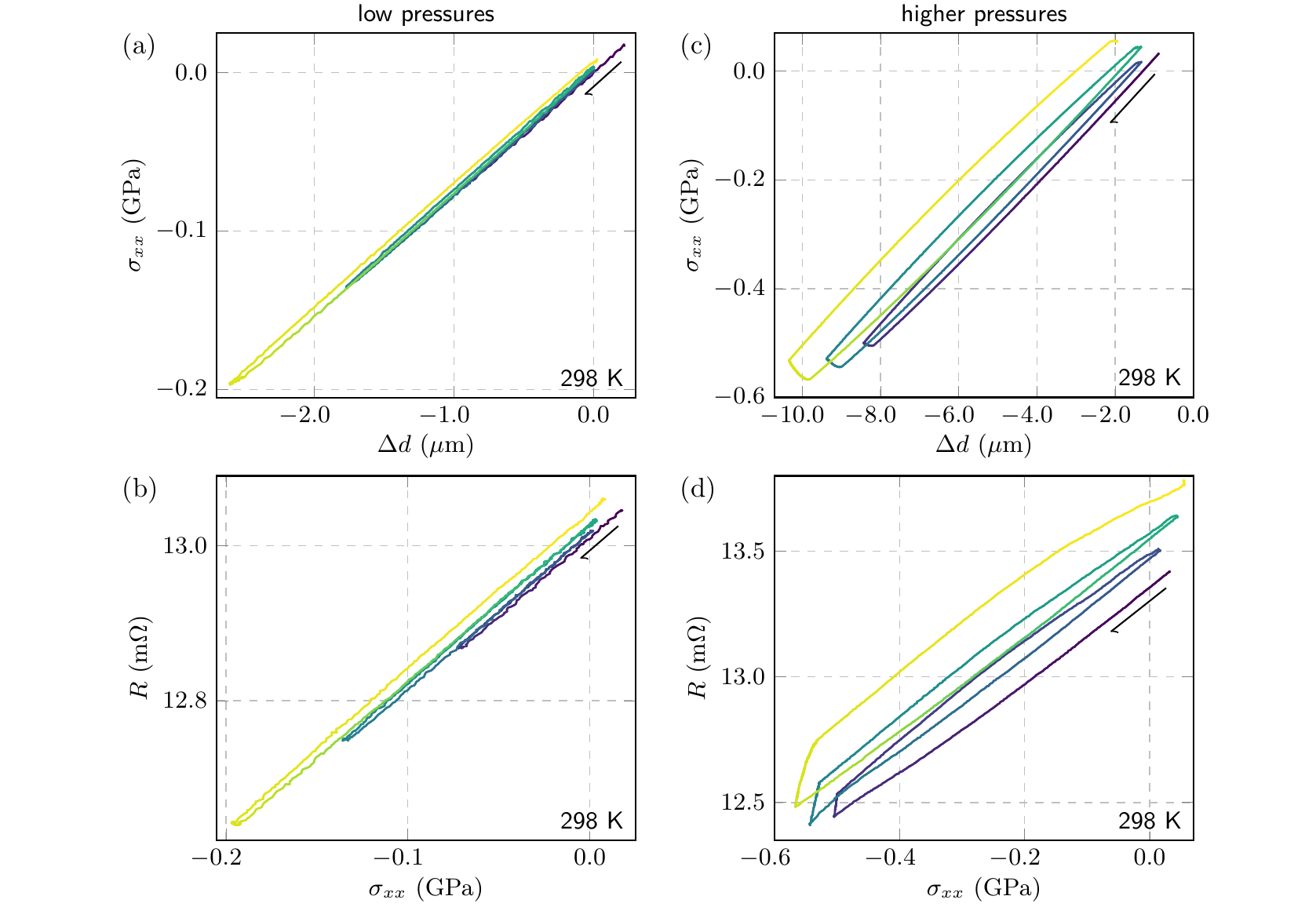}
\caption{\label{FigRT_Sr214}(a) and (c) Stress-displacement curves at room temperature for a sample of Sr$_2$RuO$_4$ covering two ranges of pressure.
(b) and (d) The sample resistance vs.\ stress during the same measurements.}
\end{figure*}

In Fig.~\ref{FigRT_Ti} we show results of measurements of a titanium sample at 298~K.
The sample comprised two pieces of titanium of cross section 310 $\times$ 70~$\mu$m, whose ends were polished flat and pressed together.
A narrow gap between the two pieces opened up as the epoxy cured; see the photograph in Fig.~\ref{FigRT_Ti}(a).
The purpose of this configuration is that when the gap is open, the force applied to the sample is definitely zero, allowing testing of the repeatability of the force sensor, and of its independence from applied displacement.

As shown in the inset of Fig.~\ref{FigRT_Ti}(a), the measured force is not fully independent of applied displacement: there is a background slope of $\sim$$20$~mN/$\mu$m.
This unwanted coupling appears because operation of the actuators causes some distortion of the outer frame, including the area around the force capacitor.

Fig.~\ref{FigRT_Ti}(a) shows the force ($F$) versus displacement ($d$) curve of this test sample at low forces, with the background slope subtracted.
Negative values of $F$ and $d$ indicate compression, and $d$ is set to zero at the point where first contact of the two titanium pieces appeared.
For this and subsequent measurements, the measurement protocol was to ramp the voltage on the compression actuator at $\sim$1~V/min.
This corresponds to a displacement ramp rate of $\sim$0.2~$\mu$m/min, although because the actuators are not linear devices the displacement ramp rate is not constant.
At the end of each ramp, the voltage was held fixed for 30 minutes.
Up to $\sim$2~N, the force-displacement curve is non-hysteretic.

Panel (b) shows measurements to higher forces, and now signs of plastic deformation appear.
In the first ramp to $|F|>2$~N, the slope of the $F(d)$ curve softens above $\approx$2.3~N.
When the actuator voltage was held fixed at the end of the ramp, $|F|$ crept to lower values while $|d|$ crept to higher values.
On the return stroke, $F(d)$ then followed a different path from the outward stroke.
From these measurements we cannot say whether the plastic deformation occurs in the sample or in the epoxy.
However it is reasonable to expect plastic deformation of the sample to dominate: the mating of the two titanium surfaces will not be perfect, resulting in stress concentration at the point(s) of initial contact.

We next turn to measurements on single crystals of the oxide metal Sr$_2$RuO$_4$ at 298~K, following the same measurement protocol.
The cross-section of the sample was 244~$\times$~119~$\mu$m, so a force of 1~N corresponds to a stress of 0.034~GPa.
The resistance of the sample was monitored during measurement to gain information on the state of the sample.

Fig.~\ref{FigRT_Sr214}(a) shows the stress-displacement curves, and panel (b) resistance versus stress, for applied uniaxial stresses $|\sigma_{xx}|$ up to 0.2~GPa.
These data suggest some plastic deformation: at higher stresses the stress-displacement curve becomes slightly nonlinear, and the sample resistance creeps upward as the stress is ramped up and down.
However the plastic deformation becomes much more obvious when the applied stress is increased further, in panels (c) and (d).
The increasing sample resistance shows that the sample deformed plastically, although it is possible that there was also plastic deformation of the epoxy holding the sample. 

\begin{figure*}[tp]
\includegraphics{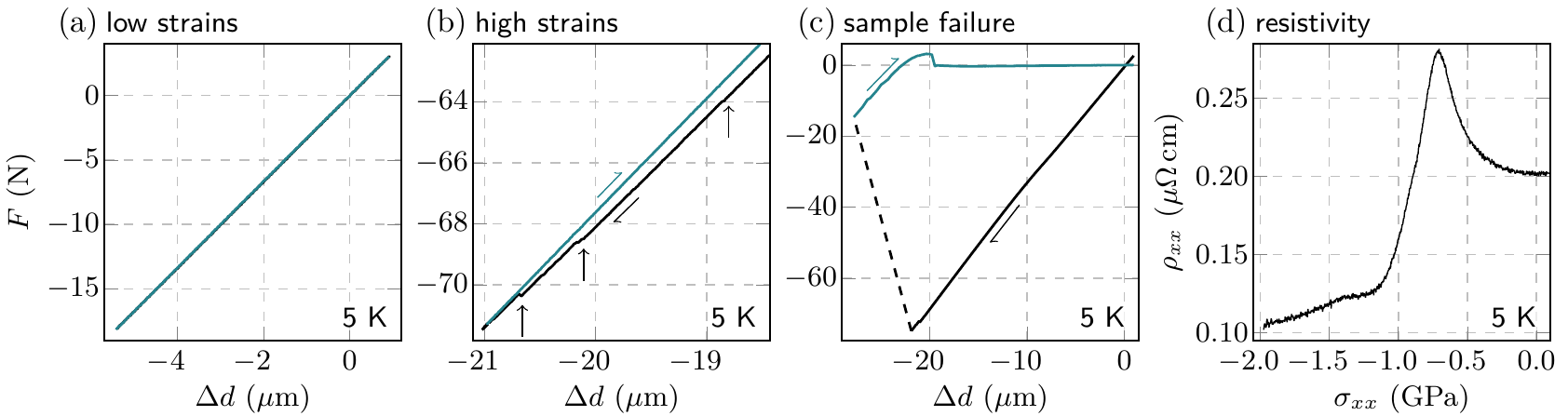}
\caption{\label{FigLT} Force-displacement curves at 5~K for a sample of Sr$_2$RuO$_4$ showing (a) the elastic regime of both the sample and the epoxy holding it, (b) minor epoxy fractures at higher applied force, marked by the vertical arrows, and (c) the sample buckling.
(d) Resistivity up to the stress where the sample buckled.}
\end{figure*}

Fig.~\ref{FigLT} shows measurements at 5~K on another sample of Sr$_2$RuO$_4$.
Because the response of the actuators is weaker at low temperatures, the voltage ramp rate was increased to 2.5~V/min.
The wait time at the end of each ramp remained 30~min.
Panel (a) shows ramps in the fully elastic regime of the sample and epoxy.
These measurements could be repeated over many cycles and reproduced the same results each time.
For these curves, the zero-stress point was taken as the stress with lowest superconducting transition temperature, in line with the previous studies on Sr$_2$RuO$_4$~\cite{Hicks2014}.
Ideally the zero-force point would be read from the force sensor.
In practice, there is variability in the force capacitor from cool-down to cool-down on a scale corresponding to a force of roughly 2~N, and improving this is a matter of better capacitor design.

At higher forces non-elastic deformation started to appear, however the form is of sudden jumps that indicate minor fracture of the epoxy and/or sample, rather than the smooth plastic deformation seen at 298~K.
A few of these jumps are highlighted in panel (b).
Each jump is to lower force and higher displacement.
The first such jump occurred at $\sim$20~N, and they continued to appear each time the strain was ramped above the previous highest point.

Fig.~\ref{FigLT}(c) shows the ramp in which the sample fractured fatally: the dashed line connects two consecutive readings of force and displacement, between which the force on the sample fell drastically while displacement increased.
The stress in the sample did not return immediately to zero because the main fracture occurred within one of the epoxy mounts.
The central portion of the sample remained in place, providing some coupling between the two mounts.

Panel (d) shows the resistivity up to the point where the sample fractured.
The large peak in resistivity at around $-$0.7~GPa is most likely the result of a Van Hove singularity in the electronic density of states being tuned to the Fermi level~\cite{Barber2018}.
This curve repeated over each successive strain ramp, both for increasing and decreasing strain, with no indications of plastic deformation in the sample.

\section{Conclusion}

A design has been presented for a piezoelectric-driven uniaxial stress cell that incorporates both force and displacement sensors.
The versatility of piezoelectric-based pressure cells has already been demonstrated, and these cells are becoming widespread.
The addition of the force sensor, however, allows more accurate and repeatable determination of the state of the sample.
This greatly enhances the usefulness of the device by allowing different samples to be compared with better quantitative precision, and by allowing early detection of nonelastic deformation.
In addition, although here we have described test samples with linear stress-strain relationships, the combination of force and displacement sensor would allow detection of stress-driven structural transitions in other materials.

The tests on Sr$_2$RuO$_4$ also highlight an important advantage of in situ application of the uniaxial stress.
At room temperature, the elastic limit of Sr$_2$RuO$_4$ was found to be $\sim$0.2~GPa.
At 5~K, in contrast, the elastic limit was at least 2.0~GPa.
In exploring the effect of uniaxial stress on electronic structure, it is therefore extremely useful to be able to cool the sample first and then apply the stress.

\begin{acknowledgments}
We acknowledge general technical discussion with Razorbill Instruments.
We thank the Max Planck Society for financial support.
Wir danken auch der Werkstatt des MPI-CPfS f\"ur die Herstellung der Bauteile.

CWH has 31\% ownership of Razorbill Instruments, a United Kingdom company that markets piezoelectric-based uniaxial stress cells.
\end{acknowledgments}

\appendix
\section{Finite element analysis of the uniaxial stress cell}

\begin{figure}[bt]
\includegraphics{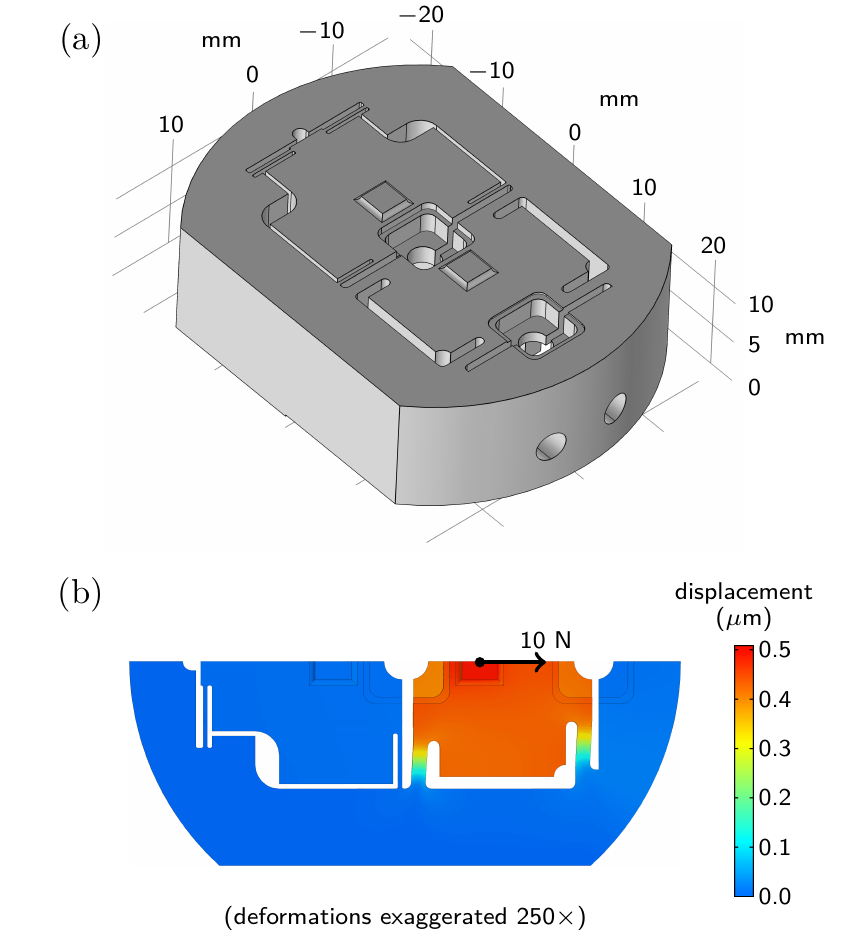}
\caption{\label{FigModel} (a) Model used for the finite element simulations and (b) the simulation used for determining the spring constant of block B. 
}
\end{figure}

We used the COMSOL Multiphysics\textsuperscript{\textregistered}\cite{COMSOL} software to perform finite element analysis (FEA) simulations of the uniaxial stress cell in operation.
Small features such as mounting holes were suppressed in the simulation model.
Two 0.5~mm-high raised platforms, 12 mm apart, which are not present in the actual cell were added to create surfaces for applying loads that simulate a sample 0.5 mm above the upper surface of the stress cell.
The model is shown in Fig.~\ref{FigModel}(a).
The material of the cell is titanium and in the FEA a Young’s modulus of 103~GPa and Poisson's ratio of 0.33 were specified.

We model the spring constant of block B by applying a load to the raised platform and measuring its displacement relative to the outer frame, which is held fixed along its two flat side faces; see Fig.~\ref{FigModel}(b).
(In the simulation we exploit the symmetry of the device and only model half of it with an imposed symmetric symmetry plane.)
The displacement of block B implies a spring constant of the set of four flexures of 20~N/$\mu$m.

Quantities that depend directly on the deformation field returned by the FEA calculations can be accurately determined with relatively simple meshes.
To study the stress or strain, which depend on the derivative of the displacement field, a mesh refinement process is required to ensure that any points of stress concentration are properly captured.
In our simulations the stress concentration occurs at the roots of the flexures and the concentration factor depends on the fillet radius.
For the simulations we use a fillet radius of 0.15~mm for the flexures guiding block A and a radius of 0.5~mm for the flexures guiding block B.
The 0.15 mm fillet approximately matches that achieved during machining, however results here should be taken as approximate.

\begin{figure}[tp]
\includegraphics{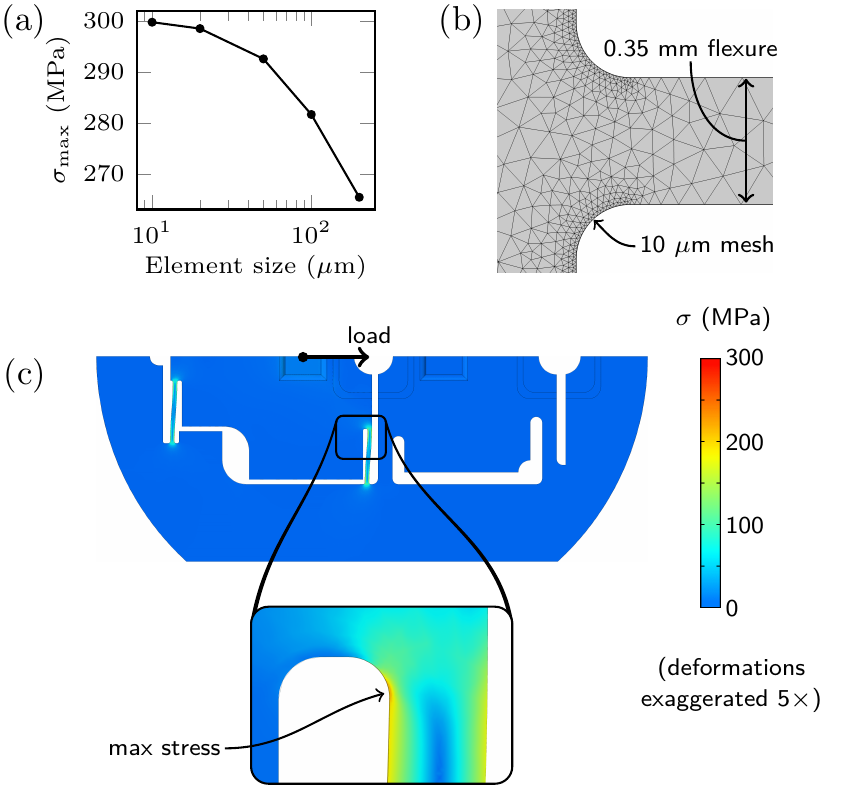}
\caption{\label{FigMaximumDisp} Finite element simulations to determine the maximum allowed displacement of block A.
}
\end{figure}

To determine the maximum displacement of block A we imposed the same fixed constraint on the two flat side faces of the body of the cell, and a load to the raised platform on block A.
The mesh density on the fillets of the flexures was increased until convergence was obtained; see Fig.~\ref{FigMaximumDisp}(a).
A maximum von Mises stress of 300~MPa was reached when the block was displaced 47~$\mu$m; see Fig.~\ref{FigMaximumDisp}(c).

\begin{figure}[tp]
\includegraphics{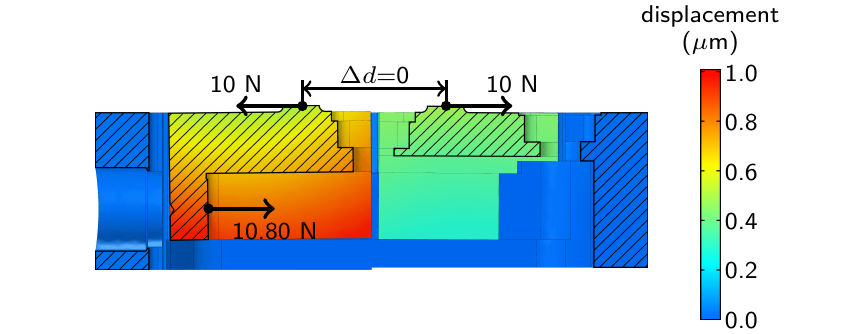}
\caption{\label{FigRotation} Finite element simulations to determine the rotational spring constant of block A and the maximum force that can be applied to a hypothetical infinitely stiff sample.
}
\end{figure}

We next determine, in a single calculation, the spring constant from rotation of block A, and the maximum stress in the in the flexures from an infinitely stiff sample.
We again make the approximation that the rotational stiffness derives completely from the flexures guiding block A, neglecting any contribution to rotational stiffness from the actuators.
To simulate application of a 10 N compressive load to an infinitely stiff sample, we apply a 10.80~N tensile load to the piezoelectric actuator attachment area of block A (which is on the bottom side of block A); see Fig.~\ref{FigRotation}, and 10.00~N load in the opposite direction to the raised platform on block A (which simulates the mounting point of the sample).
As in the previous simulations, the side faces of the pressure cell are also held fixed.
The 0.8~N load differential causes a 0.5~$\mu$m displacement of block A, as seen at the raised platform.
This 0.5~$\mu$m displacement equals the displacement expected on block B under a 10 N load, and therefore it simulates an infinitely stiff sample: the distance between the two sample attachment points does not change.
The maximum stress in the flexures after increasing the mesh density to reach convergence is 12.24~MPa.
Scaling up to a maximum allowed stress of 300~MPa implies a 245~N maximum load on the sample.

The effective spring constant for rotation of block A as seen at the sample mount point is given by the applied force divided by relative displacement between the sample mount point and the piezoelectric actuator attachment area. 
The average displacement of the piezoelectric actuator attachment area, as seen from Fig.~\ref{FigRotation}, is $\sim$0.87~$\mu$m.
The effective spring constant for rotation of block A is therefore 10~N~/~0.37~$\mu$m = 27~N/$\mu$m.

\section{Measurement of the spring constant of the uniaxial stress cell}

\begin{figure}[bt]
\includegraphics{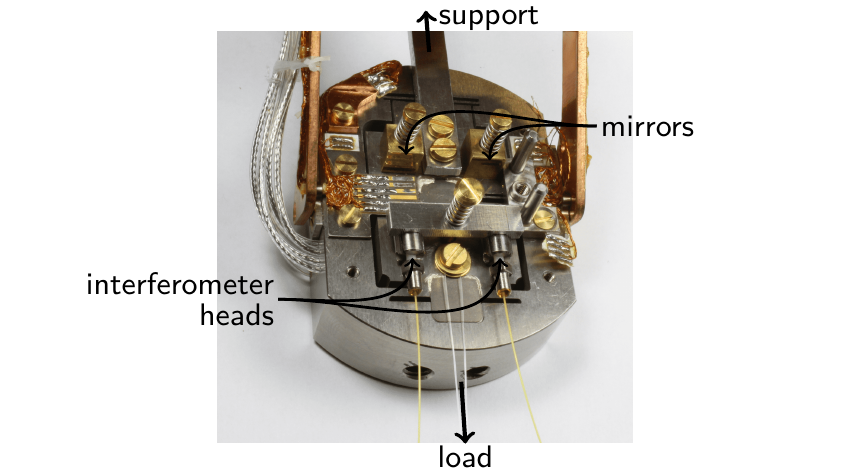}
\caption{\label{FigMeasuredSpringConst} Setup used for measuring the spring constant of the cell.
}
\end{figure}

Our setup for measuring the spring constant of the cell is shown in Fig.~\ref{FigMeasuredSpringConst}.
To measure the displacement between blocks A and B, two laser interferometer fibre heads are mounted on block B, and two mirrors on block A.
A rigid support is bolted to block A, and the load - weights hanging from a wire - is applied to block B in a way that simulates a sample mounted 0.5 mm above the surface of the cell.
Measurements with the interferometer heads 1.8 and 3.8 mm above the surface yielded springs constants of 7.5 and 7.1~N/$\mu$m, respectively, which we extrapolate to 7.8~N/$\mu$m at 0.5~mm.

\section{Calibration of the displacement and force capacitors}

The displacement-capacitance relations of the sensors were obtained using a laser interferometer.
Mirrors were attached to blocks A and B, and on the outer frame.
Three fibre heads of the interferometer were secured to the outer frame to measure the displacement of the mirrors.
The actuators were first used to ramp the displacement of block A while the interferometer displacements and the capacitance of the displacement sensor were recorded simultaneously.
This measurement was repeated in air, in vacuum at room temperature, and in vacuum at 1.5~K.
In all cases, the displacement-capacitance relation could be fitted by the parallel plate form
\[
C  = \frac{\epsilon_0 A}{ d} + C_{\text{offset}},
\]
where $C_{\text{offset}}$ accounts for stray parallel capacitances within the cell.
$A$ and $C_{\text{offset}}$ were found to be nearly identical in all three measurements.

To calibrate the force sensor, the same interferometer setup was used, and a thick titanium bar was bolted between blocks A and B to join their motion.
At low temperature, the actuators were ramped over a wide enough range to perform a fit of the parallel plate form.
At room temperature, we were concerned to avoid damage to the actuators from the high forces required to move block B, and so limited the ramp range.
Therefore we obtained only an initial value and slope $\mathrm{d}C/\mathrm{d}d$. However these values were consistent with $A$ and $C_{\text{offset}}$ remaining constant under the different measurement conditions, as for the displacement capacitor.

The spring constant of block B was determined by hanging known weights from the sample attachment point and measuring the displacement with the force capacitor.
The spring constant was found to be 19~N/$\mu$m.
The low-temperature spring constant was not measured; instead, at 5~K we use the literature value for the increase in stiffness of titanium at low temperatures, $\sim$15\%\cite{Ekin}.


\begin{thebibliography}{27}%
\makeatletter
\providecommand \@ifxundefined [1]{%
 \@ifx{#1\undefined}
}%
\providecommand \@ifnum [1]{%
 \ifnum #1\expandafter \@firstoftwo
 \else \expandafter \@secondoftwo
 \fi
}%
\providecommand \@ifx [1]{%
 \ifx #1\expandafter \@firstoftwo
 \else \expandafter \@secondoftwo
 \fi
}%
\providecommand \natexlab [1]{#1}%
\providecommand \enquote  [1]{``#1''}%
\providecommand \bibnamefont  [1]{#1}%
\providecommand \bibfnamefont [1]{#1}%
\providecommand \citenamefont [1]{#1}%
\providecommand \href@noop [0]{\@secondoftwo}%
\providecommand \href [0]{\begingroup \@sanitize@url \@href}%
\providecommand \@href[1]{\@@startlink{#1}\@@href}%
\providecommand \@@href[1]{\endgroup#1\@@endlink}%
\providecommand \@sanitize@url [0]{\catcode `\\12\catcode `\$12\catcode
  `\&12\catcode `\#12\catcode `\^12\catcode `\_12\catcode `\%12\relax}%
\providecommand \@@startlink[1]{}%
\providecommand \@@endlink[0]{}%
\providecommand \url  [0]{\begingroup\@sanitize@url \@url }%
\providecommand \@url [1]{\endgroup\@href {#1}{\urlprefix }}%
\providecommand \urlprefix  [0]{URL }%
\providecommand \Eprint [0]{\href }%
\providecommand \doibase [0]{http://dx.doi.org/}%
\providecommand \selectlanguage [0]{\@gobble}%
\providecommand \bibinfo  [0]{\@secondoftwo}%
\providecommand \bibfield  [0]{\@secondoftwo}%
\providecommand \translation [1]{[#1]}%
\providecommand \BibitemOpen [0]{}%
\providecommand \bibitemStop [0]{}%
\providecommand \bibitemNoStop [0]{.\EOS\space}%
\providecommand \EOS [0]{\spacefactor3000\relax}%
\providecommand \BibitemShut  [1]{\csname bibitem#1\endcsname}%
\let\auto@bib@innerbib\@empty
%</preamble>
\bibitem [{\citenamefont {Shirakawa}\ \emph {et~al.}(1997)\citenamefont
  {Shirakawa}, \citenamefont {Murata}, \citenamefont {Nishizaki}, \citenamefont
  {Maeno},\ and\ \citenamefont {Fujita}}]{Shirakawa1997}%
  \BibitemOpen
  \bibfield  {author} {\bibinfo {author} {\bibfnamefont {N.}~\bibnamefont
  {Shirakawa}}, \bibinfo {author} {\bibfnamefont {K.}~\bibnamefont {Murata}},
  \bibinfo {author} {\bibfnamefont {S.}~\bibnamefont {Nishizaki}}, \bibinfo
  {author} {\bibfnamefont {Y.}~\bibnamefont {Maeno}}, \ and\ \bibinfo {author}
  {\bibfnamefont {T.}~\bibnamefont {Fujita}},\ }\href {\doibase
  10.1103/PhysRevB.56.7890} {\bibfield  {journal} {\bibinfo  {journal} {Phys.
  Rev. B}\ }\textbf {\bibinfo {volume} {56}},\ \bibinfo {pages} {7890}
  (\bibinfo {year} {1997})}\BibitemShut {NoStop}%
\bibitem [{\citenamefont {Hicks}\ \emph
  {et~al.}(2014{\natexlab{a}})\citenamefont {Hicks}, \citenamefont {Brodsky},
  \citenamefont {Yelland}, \citenamefont {Gibbs}, \citenamefont {Bruin},
  \citenamefont {Barber}, \citenamefont {Edkins}, \citenamefont {Nishimura},
  \citenamefont {Yonezawa}, \citenamefont {Maeno},\ and\ \citenamefont
  {Mackenzie}}]{Hicks2014}%
  \BibitemOpen
  \bibfield  {author} {\bibinfo {author} {\bibfnamefont {C.~W.}\ \bibnamefont
  {Hicks}}, \bibinfo {author} {\bibfnamefont {D.~O.}\ \bibnamefont {Brodsky}},
  \bibinfo {author} {\bibfnamefont {E.~A.}\ \bibnamefont {Yelland}}, \bibinfo
  {author} {\bibfnamefont {A.~S.}\ \bibnamefont {Gibbs}}, \bibinfo {author}
  {\bibfnamefont {J.~A.~N.}\ \bibnamefont {Bruin}}, \bibinfo {author}
  {\bibfnamefont {M.~E.}\ \bibnamefont {Barber}}, \bibinfo {author}
  {\bibfnamefont {S.~D.}\ \bibnamefont {Edkins}}, \bibinfo {author}
  {\bibfnamefont {K.}~\bibnamefont {Nishimura}}, \bibinfo {author}
  {\bibfnamefont {S.}~\bibnamefont {Yonezawa}}, \bibinfo {author}
  {\bibfnamefont {Y.}~\bibnamefont {Maeno}}, \ and\ \bibinfo {author}
  {\bibfnamefont {A.~P.}\ \bibnamefont {Mackenzie}},\ }\href {\doibase
  10.1126/science.1248292} {\bibfield  {journal} {\bibinfo  {journal}
  {Science}\ }\textbf {\bibinfo {volume} {344}},\ \bibinfo {pages} {283}
  (\bibinfo {year} {2014}{\natexlab{a}})}\BibitemShut {NoStop}%
\bibitem [{\citenamefont {Taniguchi}\ \emph {et~al.}(2015)\citenamefont
  {Taniguchi}, \citenamefont {Nishimura}, \citenamefont {Goh}, \citenamefont
  {Yonezawa},\ and\ \citenamefont {Maeno}}]{Taniguchi2015}%
  \BibitemOpen
  \bibfield  {author} {\bibinfo {author} {\bibfnamefont {H.}~\bibnamefont
  {Taniguchi}}, \bibinfo {author} {\bibfnamefont {K.}~\bibnamefont
  {Nishimura}}, \bibinfo {author} {\bibfnamefont {S.~K.}\ \bibnamefont {Goh}},
  \bibinfo {author} {\bibfnamefont {S.}~\bibnamefont {Yonezawa}}, \ and\
  \bibinfo {author} {\bibfnamefont {Y.}~\bibnamefont {Maeno}},\ }\href
  {\doibase 10.7566/JPSJ.84.014707} {\bibfield  {journal} {\bibinfo  {journal}
  {J. Phys. Soc. Jpn.}\ }\textbf {\bibinfo {volume} {84}},\ \bibinfo {pages}
  {014707} (\bibinfo {year} {2015})}\BibitemShut {NoStop}%
\bibitem [{\citenamefont {Hicks}\ \emph
  {et~al.}(2014{\natexlab{b}})\citenamefont {Hicks}, \citenamefont {Barber},
  \citenamefont {Edkins}, \citenamefont {Brodsky},\ and\ \citenamefont
  {Mackenzie}}]{Hicks2014RSI}%
  \BibitemOpen
  \bibfield  {author} {\bibinfo {author} {\bibfnamefont {C.~W.}\ \bibnamefont
  {Hicks}}, \bibinfo {author} {\bibfnamefont {M.~E.}\ \bibnamefont {Barber}},
  \bibinfo {author} {\bibfnamefont {S.~D.}\ \bibnamefont {Edkins}}, \bibinfo
  {author} {\bibfnamefont {D.~O.}\ \bibnamefont {Brodsky}}, \ and\ \bibinfo
  {author} {\bibfnamefont {A.~P.}\ \bibnamefont {Mackenzie}},\ }\href {\doibase
  10.1063/1.4881611} {\bibfield  {journal} {\bibinfo  {journal} {Rev. Sci.
  Instrum.}\ }\textbf {\bibinfo {volume} {85}},\ \bibinfo {pages} {065003}
  (\bibinfo {year} {2014}{\natexlab{b}})}\BibitemShut {NoStop}%
\bibitem [{\citenamefont {Steppke}\ \emph {et~al.}(2017)\citenamefont
  {Steppke}, \citenamefont {Zhao}, \citenamefont {Barber}, \citenamefont
  {Scaffidi}, \citenamefont {Jerzembeck}, \citenamefont {Rosner}, \citenamefont
  {Gibbs}, \citenamefont {Maeno}, \citenamefont {Simon}, \citenamefont
  {Mackenzie},\ and\ \citenamefont {Hicks}}]{Steppke2017}%
  \BibitemOpen
  \bibfield  {author} {\bibinfo {author} {\bibfnamefont {A.}~\bibnamefont
  {Steppke}}, \bibinfo {author} {\bibfnamefont {L.}~\bibnamefont {Zhao}},
  \bibinfo {author} {\bibfnamefont {M.~E.}\ \bibnamefont {Barber}}, \bibinfo
  {author} {\bibfnamefont {T.}~\bibnamefont {Scaffidi}}, \bibinfo {author}
  {\bibfnamefont {F.}~\bibnamefont {Jerzembeck}}, \bibinfo {author}
  {\bibfnamefont {H.}~\bibnamefont {Rosner}}, \bibinfo {author} {\bibfnamefont
  {A.~S.}\ \bibnamefont {Gibbs}}, \bibinfo {author} {\bibfnamefont
  {Y.}~\bibnamefont {Maeno}}, \bibinfo {author} {\bibfnamefont {S.~H.}\
  \bibnamefont {Simon}}, \bibinfo {author} {\bibfnamefont {A.~P.}\ \bibnamefont
  {Mackenzie}}, \ and\ \bibinfo {author} {\bibfnamefont {C.~W.}\ \bibnamefont
  {Hicks}},\ }\href {\doibase 10.1126/science.aaf9398} {\bibfield  {journal}
  {\bibinfo  {journal} {Science}\ }\textbf {\bibinfo {volume} {355}},\ \bibinfo
  {pages} {eaaf9398} (\bibinfo {year} {2017})}\BibitemShut {NoStop}%
\bibitem [{\citenamefont {Barber}\ \emph {et~al.}(2018)\citenamefont {Barber},
  \citenamefont {Gibbs}, \citenamefont {Maeno}, \citenamefont {Mackenzie},\
  and\ \citenamefont {Hicks}}]{Barber2018}%
  \BibitemOpen
  \bibfield  {author} {\bibinfo {author} {\bibfnamefont {M.~E.}\ \bibnamefont
  {Barber}}, \bibinfo {author} {\bibfnamefont {A.~S.}\ \bibnamefont {Gibbs}},
  \bibinfo {author} {\bibfnamefont {Y.}~\bibnamefont {Maeno}}, \bibinfo
  {author} {\bibfnamefont {A.~P.}\ \bibnamefont {Mackenzie}}, \ and\ \bibinfo
  {author} {\bibfnamefont {C.~W.}\ \bibnamefont {Hicks}},\ }\href {\doibase
  10.1103/PhysRevLett.120.076602} {\bibfield  {journal} {\bibinfo  {journal}
  {Phys. Rev. Lett.}\ }\textbf {\bibinfo {volume} {120}},\ \bibinfo {pages}
  {076602} (\bibinfo {year} {2018})}\BibitemShut {NoStop}%
\bibitem [{\citenamefont {Stern}\ \emph {et~al.}(2017)\citenamefont {Stern},
  \citenamefont {Dzero}, \citenamefont {Galitski}, \citenamefont {Fisk},\ and\
  \citenamefont {Xia}}]{Stern2017}%
  \BibitemOpen
  \bibfield  {author} {\bibinfo {author} {\bibfnamefont {A.}~\bibnamefont
  {Stern}}, \bibinfo {author} {\bibfnamefont {M.}~\bibnamefont {Dzero}},
  \bibinfo {author} {\bibfnamefont {V.~M.}\ \bibnamefont {Galitski}}, \bibinfo
  {author} {\bibfnamefont {Z.}~\bibnamefont {Fisk}}, \ and\ \bibinfo {author}
  {\bibfnamefont {J.}~\bibnamefont {Xia}},\ }\href {\doibase 10.1038/nmat4888}
  {\bibfield  {journal} {\bibinfo  {journal} {Nat. Mater.}\ }\textbf {\bibinfo
  {volume} {16}},\ \bibinfo {pages} {708} (\bibinfo {year} {2017})}\BibitemShut
  {NoStop}%
\bibitem [{\citenamefont {Brodsky}\ \emph {et~al.}(2017)\citenamefont
  {Brodsky}, \citenamefont {Barber}, \citenamefont {Bruin}, \citenamefont
  {Borzi}, \citenamefont {Grigera}, \citenamefont {Perry}, \citenamefont
  {Mackenzie},\ and\ \citenamefont {Hicks}}]{Brodsky2017}%
  \BibitemOpen
  \bibfield  {author} {\bibinfo {author} {\bibfnamefont {D.~O.}\ \bibnamefont
  {Brodsky}}, \bibinfo {author} {\bibfnamefont {M.~E.}\ \bibnamefont {Barber}},
  \bibinfo {author} {\bibfnamefont {J.~A.~N.}\ \bibnamefont {Bruin}}, \bibinfo
  {author} {\bibfnamefont {R.~A.}\ \bibnamefont {Borzi}}, \bibinfo {author}
  {\bibfnamefont {S.~A.}\ \bibnamefont {Grigera}}, \bibinfo {author}
  {\bibfnamefont {R.~S.}\ \bibnamefont {Perry}}, \bibinfo {author}
  {\bibfnamefont {A.~P.}\ \bibnamefont {Mackenzie}}, \ and\ \bibinfo {author}
  {\bibfnamefont {C.~W.}\ \bibnamefont {Hicks}},\ }\href {\doibase
  10.1126/sciadv.1501804} {\bibfield  {journal} {\bibinfo  {journal} {Science
  Advances}\ }\textbf {\bibinfo {volume} {3}} (\bibinfo {year} {2017}),\
  10.1126/sciadv.1501804}\BibitemShut {NoStop}%
\bibitem [{\citenamefont {Zheng}\ \emph {et~al.}(2018)\citenamefont {Zheng},
  \citenamefont {Feng}, \citenamefont {Ellis},\ and\ \citenamefont
  {Kim}}]{Zheng2018}%
  \BibitemOpen
  \bibfield  {author} {\bibinfo {author} {\bibfnamefont {X.~Y.}\ \bibnamefont
  {Zheng}}, \bibinfo {author} {\bibfnamefont {R.}~\bibnamefont {Feng}},
  \bibinfo {author} {\bibfnamefont {D.~S.}\ \bibnamefont {Ellis}}, \ and\
  \bibinfo {author} {\bibfnamefont {Y.-J.}\ \bibnamefont {Kim}},\ }\href
  {\doibase 10.1063/1.5037706} {\bibfield  {journal} {\bibinfo  {journal}
  {Appl. Phys. Lett}\ }\textbf {\bibinfo {volume} {113}},\ \bibinfo {pages}
  {071906} (\bibinfo {year} {2018})}\BibitemShut {NoStop}%
\bibitem [{\citenamefont {Kim}\ \emph {et~al.}()\citenamefont {Kim},
  \citenamefont {Souliou}, \citenamefont {Barber}, \citenamefont {Lefrancois},
  \citenamefont {Minola}, \citenamefont {Tortora}, \citenamefont {Heid},
  \citenamefont {Nandi}, \citenamefont {Borzi}, \citenamefont {Garbarino},
  \citenamefont {Bosak}, \citenamefont {Porras}, \citenamefont {Loew},
  \citenamefont {K{\"o}nig}, \citenamefont {Moll}, \citenamefont {Mackenzie},
  \citenamefont {Keimer}, \citenamefont {Hicks},\ and\ \citenamefont
  {Le~Tacon}}]{Kim2018}%
  \BibitemOpen
  \bibfield  {author} {\bibinfo {author} {\bibfnamefont {H.-H.}\ \bibnamefont
  {Kim}}, \bibinfo {author} {\bibfnamefont {S.~M.}\ \bibnamefont {Souliou}},
  \bibinfo {author} {\bibfnamefont {M.~E.}\ \bibnamefont {Barber}}, \bibinfo
  {author} {\bibfnamefont {E.}~\bibnamefont {Lefrancois}}, \bibinfo {author}
  {\bibfnamefont {M.}~\bibnamefont {Minola}}, \bibinfo {author} {\bibfnamefont
  {M.}~\bibnamefont {Tortora}}, \bibinfo {author} {\bibfnamefont
  {R.}~\bibnamefont {Heid}}, \bibinfo {author} {\bibfnamefont {N.}~\bibnamefont
  {Nandi}}, \bibinfo {author} {\bibfnamefont {R.~A.}\ \bibnamefont {Borzi}},
  \bibinfo {author} {\bibfnamefont {G.}~\bibnamefont {Garbarino}}, \bibinfo
  {author} {\bibfnamefont {A.}~\bibnamefont {Bosak}}, \bibinfo {author}
  {\bibfnamefont {J.}~\bibnamefont {Porras}}, \bibinfo {author} {\bibfnamefont
  {T.}~\bibnamefont {Loew}}, \bibinfo {author} {\bibfnamefont {M.}~\bibnamefont
  {K{\"o}nig}}, \bibinfo {author} {\bibfnamefont {P.~M.}\ \bibnamefont {Moll}},
  \bibinfo {author} {\bibfnamefont {A.~P.}\ \bibnamefont {Mackenzie}}, \bibinfo
  {author} {\bibfnamefont {B.}~\bibnamefont {Keimer}}, \bibinfo {author}
  {\bibfnamefont {C.~W.}\ \bibnamefont {Hicks}}, \ and\ \bibinfo {author}
  {\bibfnamefont {M.}~\bibnamefont {Le~Tacon}},\ }\href@noop {} {\enquote
  {\bibinfo {title} {Uniaxial pressure control of competing orders in a high
  temperature superconductor},}\ }\bibinfo {note} {Submitted}\BibitemShut
  {NoStop}%
\bibitem [{\citenamefont {Kissikov}\ \emph {et~al.}(2017)\citenamefont
  {Kissikov}, \citenamefont {Sarkar}, \citenamefont {Bush}, \citenamefont
  {Lawson}, \citenamefont {Canfield},\ and\ \citenamefont
  {Curro}}]{Kissikov2017}%
  \BibitemOpen
  \bibfield  {author} {\bibinfo {author} {\bibfnamefont {T.}~\bibnamefont
  {Kissikov}}, \bibinfo {author} {\bibfnamefont {R.}~\bibnamefont {Sarkar}},
  \bibinfo {author} {\bibfnamefont {B.~T.}\ \bibnamefont {Bush}}, \bibinfo
  {author} {\bibfnamefont {M.}~\bibnamefont {Lawson}}, \bibinfo {author}
  {\bibfnamefont {P.~C.}\ \bibnamefont {Canfield}}, \ and\ \bibinfo {author}
  {\bibfnamefont {N.~J.}\ \bibnamefont {Curro}},\ }\href {\doibase
  10.1063/1.5002631} {\bibfield  {journal} {\bibinfo  {journal} {Rev. Sci.
  Instrum.}\ }\textbf {\bibinfo {volume} {88}},\ \bibinfo {pages} {103902}
  (\bibinfo {year} {2017})}\BibitemShut {NoStop}%
\bibitem [{\citenamefont {Kissikov}\ \emph {et~al.}(2018)\citenamefont
  {Kissikov}, \citenamefont {Sarkar}, \citenamefont {Lawson}, \citenamefont
  {Bush}, \citenamefont {Timmons}, \citenamefont {Tanatar}, \citenamefont
  {Prozorov}, \citenamefont {{Bud'ko}}, \citenamefont {Canfield}, \citenamefont
  {Fernandes},\ and\ \citenamefont {Curro}}]{Kissikov2018}%
  \BibitemOpen
  \bibfield  {author} {\bibinfo {author} {\bibfnamefont {T.}~\bibnamefont
  {Kissikov}}, \bibinfo {author} {\bibfnamefont {R.}~\bibnamefont {Sarkar}},
  \bibinfo {author} {\bibfnamefont {M.}~\bibnamefont {Lawson}}, \bibinfo
  {author} {\bibfnamefont {B.~T.}\ \bibnamefont {Bush}}, \bibinfo {author}
  {\bibfnamefont {E.~I.}\ \bibnamefont {Timmons}}, \bibinfo {author}
  {\bibfnamefont {M.~A.}\ \bibnamefont {Tanatar}}, \bibinfo {author}
  {\bibfnamefont {R.}~\bibnamefont {Prozorov}}, \bibinfo {author}
  {\bibfnamefont {S.~L.}\ \bibnamefont {{Bud'ko}}}, \bibinfo {author}
  {\bibfnamefont {P.~C.}\ \bibnamefont {Canfield}}, \bibinfo {author}
  {\bibfnamefont {R.~M.}\ \bibnamefont {Fernandes}}, \ and\ \bibinfo {author}
  {\bibfnamefont {N.~J.}\ \bibnamefont {Curro}},\ }\href {\doibase
  10.1038/s41467-018-03377-8} {\bibfield  {journal} {\bibinfo  {journal} {Nat.
  Commun}\ }\textbf {\bibinfo {volume} {9}},\ \bibinfo {pages} {1058} (\bibinfo
  {year} {2018})}\BibitemShut {NoStop}%
\bibitem [{\citenamefont {Watson}\ \emph {et~al.}(2018)\citenamefont {Watson},
  \citenamefont {Gibbs}, \citenamefont {Mackenzie}, \citenamefont {Hicks},\
  and\ \citenamefont {Moler}}]{Watson2018}%
  \BibitemOpen
  \bibfield  {author} {\bibinfo {author} {\bibfnamefont {C.~A.}\ \bibnamefont
  {Watson}}, \bibinfo {author} {\bibfnamefont {A.~S.}\ \bibnamefont {Gibbs}},
  \bibinfo {author} {\bibfnamefont {A.~P.}\ \bibnamefont {Mackenzie}}, \bibinfo
  {author} {\bibfnamefont {C.~W.}\ \bibnamefont {Hicks}}, \ and\ \bibinfo
  {author} {\bibfnamefont {K.~A.}\ \bibnamefont {Moler}},\ }\href {\doibase
  10.1103/PhysRevB.98.094521} {\bibfield  {journal} {\bibinfo  {journal} {Phys.
  Rev. B}\ }\textbf {\bibinfo {volume} {98}},\ \bibinfo {pages} {094521}
  (\bibinfo {year} {2018})}\BibitemShut {NoStop}%
\bibitem [{\citenamefont {Stillwell}, \citenamefont {Skove},\ and\
  \citenamefont {Davis}(1968)}]{Stillwell1968}%
  \BibitemOpen
  \bibfield  {author} {\bibinfo {author} {\bibfnamefont {E.~P.}\ \bibnamefont
  {Stillwell}}, \bibinfo {author} {\bibfnamefont {M.~J.}\ \bibnamefont
  {Skove}}, \ and\ \bibinfo {author} {\bibfnamefont {J.~H.}\ \bibnamefont
  {Davis}},\ }\href {\doibase 10.1063/1.1683303} {\bibfield  {journal}
  {\bibinfo  {journal} {Rev. Sci. Instrum.}\ }\textbf {\bibinfo {volume}
  {39}},\ \bibinfo {pages} {155} (\bibinfo {year} {1968})}\BibitemShut
  {NoStop}%
\bibitem [{\citenamefont {Overcash}\ \emph {et~al.}(1981)\citenamefont
  {Overcash}, \citenamefont {Davis}, \citenamefont {Cook},\ and\ \citenamefont
  {Skove}}]{Overcash1981}%
  \BibitemOpen
  \bibfield  {author} {\bibinfo {author} {\bibfnamefont {D.~R.}\ \bibnamefont
  {Overcash}}, \bibinfo {author} {\bibfnamefont {T.}~\bibnamefont {Davis}},
  \bibinfo {author} {\bibfnamefont {J.~W.}\ \bibnamefont {Cook}}, \ and\
  \bibinfo {author} {\bibfnamefont {M.~J.}\ \bibnamefont {Skove}},\ }\href
  {\doibase 10.1103/PhysRevLett.46.287} {\bibfield  {journal} {\bibinfo
  {journal} {Phys. Rev. Lett.}\ }\textbf {\bibinfo {volume} {46}},\ \bibinfo
  {pages} {287} (\bibinfo {year} {1981})}\BibitemShut {NoStop}%
\bibitem [{\citenamefont {Gannon}\ \emph {et~al.}(2015)\citenamefont {Gannon},
  \citenamefont {Bosak}, \citenamefont {Burkovsky}, \citenamefont {Nisbet},
  \citenamefont {Petrovi{\'c}},\ and\ \citenamefont {Hoesch}}]{Gannon2015}%
  \BibitemOpen
  \bibfield  {author} {\bibinfo {author} {\bibfnamefont {L.}~\bibnamefont
  {Gannon}}, \bibinfo {author} {\bibfnamefont {A.}~\bibnamefont {Bosak}},
  \bibinfo {author} {\bibfnamefont {R.~G.}\ \bibnamefont {Burkovsky}}, \bibinfo
  {author} {\bibfnamefont {G.}~\bibnamefont {Nisbet}}, \bibinfo {author}
  {\bibfnamefont {A.~P.}\ \bibnamefont {Petrovi{\'c}}}, \ and\ \bibinfo
  {author} {\bibfnamefont {M.}~\bibnamefont {Hoesch}},\ }\href {\doibase
  10.1063/1.4933383} {\bibfield  {journal} {\bibinfo  {journal} {Rev. Sci.
  Instrum.}\ }\textbf {\bibinfo {volume} {86}},\ \bibinfo {pages} {103904}
  (\bibinfo {year} {2015})}\BibitemShut {NoStop}%
\bibitem [{\citenamefont {Shayegan}\ \emph {et~al.}(2003)\citenamefont
  {Shayegan}, \citenamefont {Karrai}, \citenamefont {Shkolnikov}, \citenamefont
  {Vakili}, \citenamefont {De~Poortere},\ and\ \citenamefont
  {Manus}}]{Shayegan2003}%
  \BibitemOpen
  \bibfield  {author} {\bibinfo {author} {\bibfnamefont {M.}~\bibnamefont
  {Shayegan}}, \bibinfo {author} {\bibfnamefont {K.}~\bibnamefont {Karrai}},
  \bibinfo {author} {\bibfnamefont {Y.~P.}\ \bibnamefont {Shkolnikov}},
  \bibinfo {author} {\bibfnamefont {K.}~\bibnamefont {Vakili}}, \bibinfo
  {author} {\bibfnamefont {E.~P.}\ \bibnamefont {De~Poortere}}, \ and\ \bibinfo
  {author} {\bibfnamefont {S.}~\bibnamefont {Manus}},\ }\href {\doibase
  10.1063/1.1635963} {\bibfield  {journal} {\bibinfo  {journal} {Applied
  Physics Letters}\ }\textbf {\bibinfo {volume} {83}},\ \bibinfo {pages} {5235}
  (\bibinfo {year} {2003})}\BibitemShut {NoStop}%
\bibitem [{\citenamefont {Chu}\ \emph {et~al.}(2012)\citenamefont {Chu},
  \citenamefont {Kuo}, \citenamefont {Analytis},\ and\ \citenamefont
  {Fisher}}]{Chu2012}%
  \BibitemOpen
  \bibfield  {author} {\bibinfo {author} {\bibfnamefont {J.-H.}\ \bibnamefont
  {Chu}}, \bibinfo {author} {\bibfnamefont {H.-H.}\ \bibnamefont {Kuo}},
  \bibinfo {author} {\bibfnamefont {J.~G.}\ \bibnamefont {Analytis}}, \ and\
  \bibinfo {author} {\bibfnamefont {I.~R.}\ \bibnamefont {Fisher}},\ }\href
  {\doibase 10.1126/science.1221713} {\bibfield  {journal} {\bibinfo  {journal}
  {Science}\ }\textbf {\bibinfo {volume} {337}},\ \bibinfo {pages} {710}
  (\bibinfo {year} {2012})}\BibitemShut {NoStop}%
\bibitem [{PIC(2016)}]{PICatalog}%
  \BibitemOpen
  \href@noop {} {\emph {\bibinfo {title} {Catalog: Piezoelectric Ceramic
  Products: Fundamentals, Characteristics and Applications}}},\ \bibinfo
  {organization} {PI Ceramic GmbH} (\bibinfo {year} {2016})\BibitemShut
  {NoStop}%
\bibitem [{PIC()}]{PICMA}%
  \BibitemOpen
  \href@noop {} {}\bibinfo {note} {Physik Instrumente P-885.51
  {PICMA\textsuperscript{\textregistered}} Stack Multilayer Piezo
  Actuator}\BibitemShut {NoStop}%
\bibitem [{PIU(2017)}]{PIUserManual}%
  \BibitemOpen
  \href@noop {} {\emph {\bibinfo {title} {PZ259E P-­88x/P-­08x Piezo Actuator
  User Manual}}},\ \bibinfo {organization} {PI Ceramic GmbH} (\bibinfo {year}
  {2017})\BibitemShut {NoStop}%
\bibitem [{\citenamefont {Leyens}\ and\ \citenamefont
  {Peters}(2003)}]{Leyens2003}%
  \BibitemOpen
  \bibfield  {author} {\bibinfo {author} {\bibfnamefont {C.}~\bibnamefont
  {Leyens}}\ and\ \bibinfo {author} {\bibfnamefont {M.}~\bibnamefont
  {Peters}},\ }\href {\doibase 10.1002/3527602119} {\emph {\bibinfo {title}
  {Titanium and Titanium Alloys: Fundamentals and Applications}}}\ (\bibinfo
  {publisher} {Wiley},\ \bibinfo {year} {2003})\BibitemShut {NoStop}%
\bibitem [{\citenamefont {Barber}(2017)}]{Barber2017}%
  \BibitemOpen
  \bibfield  {author} {\bibinfo {author} {\bibfnamefont {M.~E.}\ \bibnamefont
  {Barber}},\ }\href {http://hdl.handle.net/10023/15429} {\bibinfo {type}
  {{Ph.D.} thesis}},\ \bibinfo  {school} {University of St Andrews} (\bibinfo
  {year} {2017})\BibitemShut {NoStop}%
\bibitem [{Note1()}]{Note1}%
  \BibitemOpen
  \bibinfo {note} {The elastic properties of Stycast 2850FT with Catalyst 23LV
  have not been measured at very low temperatures. By comparing the measured
  Young's modulus with those from other Catalysts that have been extended to
  lower temperatures and with the unfilled version of Stycast, 1266, we
  estimate a Young's modulus of 15~GPa for Stycast 2850FT with Catalyst 23LV at
  low temperatures. The shear modulus of an isotropic material is $G = E / 2(1
  + \nu )$, where $\nu $ is Poisson's ratio. We take $\nu \sim 0.3$, yielding
  $G \sim 6$~GPa. Further details are given in Ref.~\protect \rev@citealpnum
  {Hicks2014RSI}}\BibitemShut {NoStop}%
\bibitem [{\citenamefont {Paglione}\ \emph {et~al.}(2002)\citenamefont
  {Paglione}, \citenamefont {Lupien}, \citenamefont {MacFarlane}, \citenamefont
  {Perz}, \citenamefont {Taillefer}, \citenamefont {Mao},\ and\ \citenamefont
  {Maeno}}]{Paglione2002}%
  \BibitemOpen
  \bibfield  {author} {\bibinfo {author} {\bibfnamefont {J.}~\bibnamefont
  {Paglione}}, \bibinfo {author} {\bibfnamefont {C.}~\bibnamefont {Lupien}},
  \bibinfo {author} {\bibfnamefont {W.~A.}\ \bibnamefont {MacFarlane}},
  \bibinfo {author} {\bibfnamefont {J.~M.}\ \bibnamefont {Perz}}, \bibinfo
  {author} {\bibfnamefont {L.}~\bibnamefont {Taillefer}}, \bibinfo {author}
  {\bibfnamefont {Z.~Q.}\ \bibnamefont {Mao}}, \ and\ \bibinfo {author}
  {\bibfnamefont {Y.}~\bibnamefont {Maeno}},\ }\href {\doibase
  10.1103/PhysRevB.65.220506} {\bibfield  {journal} {\bibinfo  {journal} {Phys.
  Rev. B}\ }\textbf {\bibinfo {volume} {65}},\ \bibinfo {pages} {220506}
  (\bibinfo {year} {2002})}\BibitemShut {NoStop}%
\bibitem [{COM()}]{COMSOL}%
  \BibitemOpen
  \href@noop {} {}\bibinfo {note} {{COMSOL
  Multiphysics{\textsuperscript{\textregistered}} v.\ 5.3a.
  \url{www.comsol.com}. COMSOL AB, Stockholm, Sweden}}\BibitemShut {NoStop}%
\bibitem [{\citenamefont {Ekin}(2006)}]{Ekin}%
  \BibitemOpen
  \bibfield  {author} {\bibinfo {author} {\bibfnamefont {J.}~\bibnamefont
  {Ekin}},\ }\href {\doibase 10.1093/acprof:oso/9780198570547.001.0001} {\emph
  {\bibinfo {title} {Experimental Techniques for Low-Temperature Measurements:
  Cryostat Design, Material Properties and Superconductor Critical-Current
  Testing}}}\ (\bibinfo  {publisher} {Oxford University Press},\ \bibinfo
  {year} {2006})\BibitemShut {NoStop}%
\end{thebibliography}
\end{document}